\documentclass[superscriptaddress,amsmath,amssymb,aps,prb,floatfix,twocolumn,showpacs]{revtex4}
\usepackage{graphicx}
\usepackage{bm}
\newcommand{\buck}{{C$_{60}$ }}
\newcommand{\beq}{\begin{equation}}
\newcommand{\eeq}{\end{equation}}

\newcommand{\beqa}{\begin{eqnarray}}
\newcommand{\eeqa}{\end{eqnarray}}

\begin{document}

\title{Light emission and finite frequency shot noise in molecular junctions: from tunneling to contact}
\author{Jing-Tao \surname{L\"u}}
\email{jtlu@hust.edu.cn}
\affiliation{School of Physics, Huazhong University of Science and Technology, Wuhan, China}
\affiliation{DTU-Nanotech, Department of Micro- and Nanotechnology, Technical
University of Denmark, {\O}rsteds Plads, Bldg.~345E, DK-2800 Kongens
Lyngby, Denmark}
\affiliation{Niels-Bohr Institute, Nano-Science Center, University of
Copenhagen, Universitetsparken 5, 2100 Copenhagen {\O}, Denmark}
\author{Rasmus Bjerregaard Christensen}
\affiliation{DTU-Nanotech, Department of Micro- and Nanotechnology, Technical
University of Denmark, {\O}rsteds Plads, Bldg.~345E, DK-2800 Kongens
Lyngby, Denmark}
\author{Mads \surname{Brandbyge}}
\affiliation{DTU-Nanotech, Department of Micro- and Nanotechnology, Technical
University of Denmark, {\O}rsteds Plads, Bldg.~345E, DK-2800 Kongens
Lyngby, Denmark}

\begin{abstract} 
Scanning tunneling microscope induced light emission from an atomic or
molecular junction has been probed from the tunneling to contact regime in
recent experiments.  There, the measured light emission yields suggest a strong
correlation with the high frequency current/charge fluctuations. We show that
this is consistent with the established theory in the tunneling regime, by
writing the finite-frequency shot noise as a sum of inelastic transitions
between different electronic states. Based on this, we develop a practical
scheme to perform calculations on realistic structures using nonequilibrium Green's functions.
The photon emission yields obtained re-produce the essential feature of the
experiments.  
\end{abstract}
\maketitle 
\section{Introduction}
When a scanning tunneling microscope (STM) tip is brought towards a metal
surface, strong localized plasmon modes develop between the tip and
surface, in addition to the propagating surface mode at the metal interface.
Under an electric field, the plasmon modes interact with the electrons traversing the
gap. This provides an efficient way to excite the plasmon modes electrically,
and has become an important topic bridging nanoelectronics and
plasmonics\cite{NiGa.12,chen2009,ward2010,Kern2012,GiSaScSc89,Qiu2003, Dong2004,
Cavar2005, Uemura2007, Marquardt2010, Hoffmann2002,Tao2009,Schneider2011,Berndt1991,Berndt1993,Berndt1993a,Aizpurua2000,Zhang2009,Geng2012,WaBoZhCoDu11,BhBoNo11}. Radiative damping of the excited plasmons results in light emission, which can be detected experimentally in the far field at the
same or opposite side of the STM tip\cite{GiSaScSc89,Qiu2003, Dong2004,
Cavar2005, Uemura2007, Marquardt2010, Hoffmann2002,
Tao2009,Schneider2011,WaBoZhCoDu11}. Analyzing the emitted light can provide information about the nanogap. The dependence of light emission on
the type of metal, the shape of tip and surface, and on the inserted molecular
layer between tip and surface, have all been explored\cite{Berndt1991,Berndt1993,Berndt1993a,Aizpurua2000,Zhang2009,Geng2012,LuGrDe.13}.
Different types of plasmon modes have been
detected\cite{WaBoZhCoDu11,BhBoNo11}. Most of these experiments are done in the
tunneling regime, where the coupling between STM tip and metal surface is weak.
Theoretically, it has been established that the excitation of plasmon modes is
due to the inelastic electronic transitions taken place near the gap\cite{JoMoAp90,PERSSON1992}. 

Recently, STM-induced light emission has been probed during the transition from
the tunneling to the contact regime, both for single atom contacts and a \buck
molecular junction\cite{ScNeJo.2009,ScScBe10,ScLuBrBe12}. The experimental
results suggest a strong correlation between the light emission intensity and
the current/charge fluctuations at optical frequencies, and furthermore, show
the possibility of controlling light emission by engineering the electronic
structure. The established theory in the weak coupling, tunnelling regime seems 
to be inadequate for explaining the experimental results in the strong coupling, contact regime.

A detailed modeling of such experiments needs to take into account the plasmon
field distribution near the STM tip, the nonequilibrium electronic structure at
high bias, the coupling of the plasmonic field with electrical current, and the
propagation of light to the far field\cite{JoMoAp90,PERSSON1992,marty2012}.  In
this paper, instead of developing a full theory, we focus on the electronic
part of the problem. In particular, we study how the change of the electronic
structure with tip-position and voltage bias influences the efficiency of plasmonic excitation.  
To this end,  we derive a Fermi-golden-rule like expression for the finite frequency shot noise,
and relate it to the theory of STM-induced light emission in the tunneling
regime. We then express the result in terms of nonequilibrium Green's functions
(NEGF), and develop a practical scheme to perform calculations on realistic
structures, using information available from Density Functional Theory based
NEGF (DFT-NEGF) transport calculations.  We demonstrate how this scheme manage
to capture the essential feature of the atomic metal and molecular contact
experiments.

\section{Theory}
In this section, we briefly summarize the theory of STM-induced light emission
in the tunneling regime\cite{JoMoAp90,PERSSON1992}. Then, following
Ref.~\onlinecite{Gavish2000}-\onlinecite{AgKo00}, we introduce an approach to express the finite
frequency shot noise in a coherent conductor as a sum of inelastic
electronic transitions. 
We demonstrate how the shot-noise explanation of the
light emission in a molecular contact is consistent with the theory in the
tunneling regime.

\subsection{Inelastic transition due to electron-plasmon interaction}
Following the theory of light emission from STM\cite{JoMoAp90,PERSSON1992} and point
contacts\cite{LeLeBl.2010}, the interaction of the electrical current with the
plasmon field in the tip-surface cavity is described by the following
Hamiltonian,
\begin{equation}
	{H}_{int}=\frac{1}{c}\int {j}(r){A}(r)d^3 r\,,
	\label{eq:epi}
\end{equation}
where ${j}(r)$ is the electron current density operator at position $r$. The plasmon
mode, with frequency, $\Omega$, and spatial distribution, $\xi(r)$, is represented by a vector potential,
\begin{equation}
	{A}(r)= \sqrt{\frac{2\pi\hbar c^2}{V \Omega}}\xi(r) \left(a + a^\dagger \right)\,.
	\label{eq:plas}
\end{equation}
Here $a$($a^\dagger$) is the annihilation (creation) operator of the plasmon
mode, $c$ is the speed of the light, $\hbar$ the reduced Planck constant, and
$V$ the normalization volume. In principle, we may calculate the plasmon mode
frequency and field distribution for a given a tip-surface distance. However,
this is a daunting task for atomistic first principles theory and we do not
consider this problem here.  Instead, we focus only on the source of the light
emission, and investigate the effect of the non-equilibrium electronic
structure on the emission rate. We ignore the spatial distribution of the mode
in the $xy$-plane transverse to the current, $\xi(r) = \xi(z)$, and perform the
integration over these directions in Eq.~(\ref{eq:epi}) and get
\begin{eqnarray}
	{H}_{int}&=&\frac{1}{c}\int {I}(z){A}(z)d z,\nonumber\\
	&=& M (a+a^\dagger) ,
	\label{eq:epi2}
\end{eqnarray}
where ${I}(z)$ is the surface current evaluated at $z$, integrated over the
transverse surface. The emitted power from the junction is proportional to the
inelastic transition probability due to the interaction between initial($\psi_i$) and final($\psi_f$) states originating from the tip or surface electrode,
\begin{eqnarray}
	\label{eq:fgr}
	P(\Omega) \!&\sim&\! \sum_{i,f}\!\int\!\!\!  \int\! |\langle \psi_f|M|\psi_i\rangle|^2 \delta(\varepsilon_i-\varepsilon_f-\hbar\Omega)\\
	&&\times n_F(\varepsilon_i-\mu_i)(1-n_F(\varepsilon_f-\mu_f)) d\varepsilon_i d\varepsilon_f\,.\nonumber
\end{eqnarray}
We employ the normalization, $\langle \psi_i| \psi_j\rangle =
\delta_{ij} \delta(\varepsilon_i-\varepsilon_j)$,
and filling given by the Fermi-Dirac distributions, $n_F$, corresponding to the initial and final electrodes with Fermi energies given by
$\mu_i$ and $\mu_f$, respectively. Finally, we will assume that the ``diagonal''
contributions in the $z$ direction capture the main dependence of the emitted power on the electronic structure of the junction. Thus we get,
\begin{eqnarray}
	\label{eq:fgr2}
	P(\Omega) &\sim&  \int dz\: |\xi(z)|^2 \sum_{i,f}\int\!\!\! \int |\langle \psi_f|{I}(z)|\psi_i\rangle|^2\delta(\varepsilon_i-\varepsilon_f-\hbar\Omega)\nonumber\\
	&&\times n_F(\varepsilon_i-\mu_i)(1-n_F(\varepsilon_f-\mu_f)) d\varepsilon_i d\varepsilon_f.
\end{eqnarray}
This "diagonal" assumption can clearly not be justified {\it per se} without
concrete knowledge about the spatial distribution of the mode along with the
local current operator. However, below we will use a first principles method in
order to calculate without any fitting parameters the light emission using this
approximation and compare with the experimental trends.

\subsection{Current, charge fluctuations and emission rate}
Now we show that the Fermi's golden-rule rate in Eq.~(\ref{eq:fgr2}) is closely
related to the finite frequency shot noise of the electrical current, which is
defined as
\begin{equation}
	\langle\langle {I}_z(0) {I}_{z'}(t)\rangle\rangle \equiv \langle ({I}_z(0)-\langle {I}_z(0)\rangle)({I}_{z'}(t)-\langle {I}_{z'}(t)\rangle)\rangle,
\label{eq:cn}
\end{equation}
where ${I}(t)=e^{iHt/\hbar}{I}e^{-iHt/\hbar}$ is the surface current operator along $z$ 
in the Heisenberg representation and $z$/${z'}$ are two positions along the
transport direction. The positive direction of $I_z$ is defined to be from
the surface electrode towards the tip. Since we are dealing with the time
dependence explicitly, we put the position variables $z$, $z'$ as the
sub-indices. The Fourier transform of Eq.~(\ref{eq:cn}) gives the noise
spectrum,
\begin{equation}
	S_{z{z'}}(\omega) = \int_{-\infty}^{+\infty} \langle\langle {I}_z(0) {I}_{z'}(t)\rangle\rangle e^{i\omega t}dt\,.
	\label{eq:fourier}
\end{equation}
Following Ref.~\onlinecite{Gavish2000}-\onlinecite{AgKo00}, inserting a complete set of eigenstates
into Eq.~(\ref{eq:fourier}), and doing the Fourier transform, we obtain a golden-rule-type expression for the current noise,
\begin{eqnarray}
	S_{z{z'}}(\omega) &=& 2\pi\hbar\sum_{\begin{subarray}{l}i,f\\i\neq f\end{subarray}}\int\!\!\!\!\int\langle\psi_{i}|{I}_{z}|\psi_{f}\rangle\langle\psi_{f}|{I}_{{z'}}|\psi_{i}\rangle\delta(\varepsilon_i-\varepsilon_f-\hbar\omega)\nonumber\\
 &&\times  n_F(\varepsilon_i-\mu_i)(1-n_F(\varepsilon_f-\mu_f))d\varepsilon_i d\varepsilon_f\,.
	\label{eq:tn}
\end{eqnarray}
The initial and final states are summed over scattering states from both
electrodes.  Equation~(\ref{eq:tn}) includes both the Nyquist-Johnson (thermal)
and shot noise contributions. Since the energy of the emitted light is much
larger than the thermal energy ($\hbar\omega\gg k_B T$), only the
zero-temperature limit is considered. In this case, besides the zero-point
fluctuations, the only contribution is the shot noise,
\begin{eqnarray}
\label{eq:rate}
	S_{z{z'}}(\omega) &=&2\pi\hbar\sum_{s,t}\int_{\mu_s+\hbar\omega}^{\mu_t}\langle\psi_{t}|{I}_{z}|\psi_{s}\rangle\langle\psi_{s}|{I}_{{z'}}|\psi_{t}\rangle d\varepsilon_t\,,\nonumber\\
\end{eqnarray}
with $\varepsilon_s=\varepsilon_t-\hbar\omega$ for positive sample bias $V=V_s-V_t>0$. We define the upper and lower Fermi levels are at $|eV|/2$ and $-|eV|/2$, respectively. 
The ''diagonal'' correlation $S_{zz}$ gives the sum of the transition rates
between the initial filled tip scattering states $\psi_t$, and the final empty
surface scattering states $\psi_s$, with energies $\varepsilon_t$ and
$\varepsilon_s$, respectively. This illustrates how the finite frequency shot noise can be viewed as inelastic electronic transitions between the tip and surface scattering
states. The positive frequency/energy part of the noise spectrum corresponds to
the photon emission, relevant to the experiment, and the negative part to the
absorption process. We notice that if $z$ and $z'$ are located at the surface and tip
electrode, respectively, then according to charge conservation, 
\begin{equation}
I_{d} \equiv \dot Q_d = I_{z}-I_{z'}\,,
\label{eq:sdd0}
\end{equation}
and therefore, the charge fluctuation in the central molecule/"device" region($d$) is given by:
\begin{equation}
S_{dd} = S_{zz}+S_{z'z'}-S_{zz'}-S_{z'z}\,.
\label{eq:sdd}
\end{equation}
Similarly the fluctuation of the average current $I_a=\frac{1}{2}(I_z+I_{z'})$ is:
\begin{equation}
	S_{aa} = \frac{1}{4}\left(S_{zz}+S_{z'z'}+S_{zz'}+S_{z'z}\right).
\label{eq:saa}
\end{equation}

Using the result in this subsection, we can write Eq.~(\ref{eq:fgr2}) as
\begin{eqnarray}
	\label{eq:ratefinal}
	P(\Omega) &\sim&  \int dz\: |\xi(z)|^2 S_{zz}(\Omega),
\end{eqnarray}
which makes connection between the `old' theory for STM-induced light emission
in the tunneling regime and the `new' shot noise argument.

\section{Numerical scheme}
We aim at a formulation targeting the DFT-NEGF approach to atomistic electron
transport,  such as the SIESTA/TranSIESTA method\cite{BrMoOr.02} and similar
methods employing a localized basis set.  In these the whole system is
separated into a central device region($d$), and two electrode regions, here
the tip ($t$) and surface ($s$) electrodes.  The electrodes are represented by
the self-energies. In order to directly employ the DFT-NEGF formalism we will
rewrite Eq.~(\ref{eq:rate}) in terms of the device Green's functions and the
self-energies ($\Sigma_s$,$\Sigma_t$) folded into the same device region
representing the coupling of the device region to tip and surface electrodes,
respectively. By our choice of device region we effectively define separating
surfaces between the regions. 

As an example we now consider the current evaluated at the surface electrode.
In order to calculate the surface electrode current fluctuations,
$S_{ss}(\omega)$, an explicit expression for the surface current is needed in
terms of quantities readily available in the DFT-NEGF calculation. The current
matrix ${I}_s$, can be written as\cite{PaBr07},
\begin{equation}
	I_s = -\frac{ie}{\hbar}[P_s,H]=\frac{ie}{\hbar}(V_{ds}-{V}_{sd}),
	\label{eq:is}
\end{equation}
where $P_s$ denotes projection into the surface electrode subspace, $H$ is the
total Hamiltonian, $V_{ds}$ is the coupling matrix between the device and
surface electrode, $V_{sd}$ is its complex conjugate, and $e$ is the electron
charge. We ignore electron spin throughout the paper, since it is not relevant.
We assume an orthogonal basis set; however, a generalization to the
non-orthogonal case is straightforward by a L\"owdin transformation.  

Next, we evaluate the current matrix element between different scattering
states. We start from the Lippmann-Schwinger equation connecting the scattering
states and the retarded Green's functions of the whole system $G(\varepsilon)$,
\begin{eqnarray}
	|\psi_s(\varepsilon)\rangle = |\phi_s(\varepsilon)\rangle  + G(\varepsilon) V_{T} |\phi_s(\varepsilon)\rangle\,.
	\label{eq:ls}
\end{eqnarray}
Here $|\psi_s(\varepsilon)\rangle$ and $|\phi_s(\varepsilon)\rangle$ are the
scattering states from the semi-infinite surface electrode with and without
coupling to the device, respectively.  Note that $\phi_s$ is non-zero only in
the surface electrode, but $\psi_s$ spans over the whole region including both
electrodes and the device. The coupling matrix, $V_T$, represent the coupling
between the device and the two electrodes, localized near the device-electrode
interfaces. Here $G(\varepsilon)$ is the retarded Green's function of the whole
system including the effect of $V_T$.

Using the projection matrices, $P_t + P_d+P_s=I$, and the fact that
$V_T|\phi_s\rangle$ is only non-zero in the device region, it is possible to write the current matrix element $\langle
\psi_{t}(\varepsilon)|{I}_s|\psi_{s}(\varepsilon_-)\rangle$ in terms of the
device Green's functions and self-energies, where $\varepsilon_-=\varepsilon-\hbar\omega$. 
Firstly, using $V_{ds}=P_dV_{ds}P_s$,
and Eq.~(\ref{eq:ls}), we have
\begin{equation}
	P_s|\psi_{s}(\varepsilon_-)\rangle =(I+G_{sd}(\varepsilon_-)V_{ds})|\phi_s(\varepsilon_-)\rangle.
	\label{}
\end{equation}
Here $G_{sd} \equiv P_s G P_d$ is a submatrix of the full Green's function $G$,
 and $G_{dd}$ is defined correspondingly. Using the relations, 
\begin{eqnarray}
	G_{sd} &=& g_{ss}V_{sd}G_{dd},\\
	|\psi_{s}^{d}\rangle &=& P_d|\psi_{s}\rangle = G_{dd}V_{ds}|\phi_s\rangle,\\
	\Sigma_{s} &=& V_{ds} g_{ss}V_{sd},
	\label{}
\end{eqnarray}
we get,
\begin{equation}
	\langle \psi_{t}(\varepsilon)|V_{ds}|\psi_{s}(\varepsilon_-)\rangle = \langle \psi^d_{t}(\varepsilon)|G_{dd}^{-1}(\varepsilon_-)+\Sigma_{s}(\varepsilon_-)|\psi^d_{s}(\varepsilon_-)\rangle.
	\label{}
\end{equation}
Note that here $g_{ss}$ is the retarded Green's function of the isolated surface electrode.
Similarly, for the second term in Eq.~(\ref{eq:is}), we have
\begin{eqnarray}
	\langle \psi_{t}(\varepsilon)|V_{sd}|\psi_{s}(\varepsilon_-)\rangle &=& \langle \psi_{t}(\varepsilon)|P_sV_{sd}P_d|\psi_{s}(\varepsilon_-)\rangle \nonumber\\
	&=& \langle \psi_{t}(\varepsilon)|V_{td}G^\dagger_{dd}V_{ds}G^\dagger_{ss}V_{sd}P_d|\psi_{s}(\varepsilon_-)\rangle \nonumber\\
	&=& \langle \psi^d_{t}(\varepsilon)|\Sigma_s^\dagger(\varepsilon) |\psi^d_{s}(\varepsilon_-)\rangle.
	\label{}
\end{eqnarray}
Defining
\begin{eqnarray}
W_i(\varepsilon_-,\varepsilon)\equiv G_d^{-1}(\varepsilon_-)+\Sigma_i(\varepsilon_-)-\Sigma_i^\dagger(\varepsilon),
\end{eqnarray}
we finally obtain the desired matrixelement,
\begin{eqnarray}
	\langle \psi_{t}(\varepsilon)|{I}_s|\psi_{s}(\varepsilon_-)\rangle &=& \frac{ie}{\hbar} \langle \psi^d_{t}(\varepsilon)|W_s(\varepsilon_-,\varepsilon)|\psi^d_{s}(\varepsilon_-)\rangle\,.\nonumber\\
\end{eqnarray}
Note that all quantities are projected to the device region and thus depend on the actual splitting into regions.

Using the current matrix element, we can now write the surface current shot noise at zero
temperature as,
\begin{eqnarray}
	\label{eq:ratenegf}
	S_{ss}(\omega)\!\! &=&\!\! \int_{\theta}{\rm Tr}\left[ W_s(\varepsilon_-,\varepsilon)A_{s}(\varepsilon_-)W_s^\dagger(\varepsilon_-,\varepsilon) A_{t}(\varepsilon)\right] d\varepsilon\,,\nonumber\\
\end{eqnarray}
where the integral is defined as,
\begin{equation}
	\int_\theta \cdot \:\:d\varepsilon = \theta(|eV|-\hbar\omega)\frac{e^2}{2\pi\hbar}\int_{\hbar\omega-|eV|/2}^{|eV|/2} \cdot\:\: d\varepsilon,
	\label{}
\end{equation}
with $\theta(x)$ being the Heaviside step function, $A_s(\varepsilon) =
G_d(\varepsilon)\Gamma_s(\varepsilon)G_d^\dagger(\varepsilon) = 2\pi
\sum_{i=s}|\psi^d_i(\varepsilon)\rangle\langle\psi^d_i(\varepsilon)|$ is the device
spectral function due to scattering states from the surface electrode,
similarly for $A_t$, and $\Gamma_s=i(\Sigma_s-\Sigma_s^\dagger)$.
In the same way, we get the tip current noise,
\begin{eqnarray}
	\label{eq:ratenegf2}
	S_{tt}(\omega)\!\!\! &=& \!\!\!\int_{\theta}^{}{\rm Tr}\!\left[ W^\dagger_t(\varepsilon,\varepsilon_-)A_{s}(\varepsilon_-)W_t(\varepsilon,\varepsilon_-) A_{t}(\varepsilon)\right] d\varepsilon\,,\nonumber\\
\end{eqnarray}
and their cross correlation,
\begin{eqnarray}
	\label{eq:ratenegf3}
	S_{st}(\omega)&=&S_{ts}^*(\omega)\\
	&=& -\int_{\theta}^{}{\rm Tr}\left[ W_s(\varepsilon_-,\varepsilon)A_{s}(\varepsilon_-)W_t(\varepsilon,\varepsilon_-) A_{t}(\varepsilon)\right] d\varepsilon\,.\nonumber
\end{eqnarray}
Equations~(\ref{eq:ratenegf}-\ref{eq:ratenegf3}) are our main formal results, where we
have written the finite frequency shot noise in terms of the Green's functions
and self-energies, readily available from DFT-NEGF calculations. 
The difference between Eqs.~(\ref{eq:ratenegf}) and ~(\ref{eq:ratenegf2}) reveals
the position dependence of finite frequency noise. Importantly, they both yield the
standard result in the zero-frequency limit\cite{BlBu.2000}. 

Assuming constant self-energies ($\Sigma_s,\Sigma_t$), and decoupled
eigenchannel transmissions\cite{PaBr07} at different energies, $T_n(\varepsilon)$, we arrive at more physically transparent
expressions,
\begin{eqnarray}
S_{ss}(\omega)&=&\sum_{n}\int_{\theta}^{}\: T_{n}(\varepsilon)(1-T_{n}(\varepsilon_-))\: d\varepsilon\,,\label{eq:arate1-1}\\
S_{tt}(\omega)&=&\sum_{n}\int_{\theta}^{}\: T_{n}(\varepsilon_-)(1-T_{n}(\varepsilon))\: d\varepsilon\,, \label{eq:arate1-2}
\end{eqnarray}
valid for positive sample voltages, $V>0$.  The two expressions are exchanged for negative bias.
Note that $T_n$ are the channel transmissions calculated for the particular bias, $V$. 
We refer to appendix \ref{app.A} for the full result of $S_{ss}(\omega)$ at finite temperature.   
Unfortunately, we are not able to write the cross correlations $S_{st}$
and $S_{ts}$ in terms of the eigentransmissions $T_n$.

\begin{figure}[htbp]
   \includegraphics[scale=0.3]{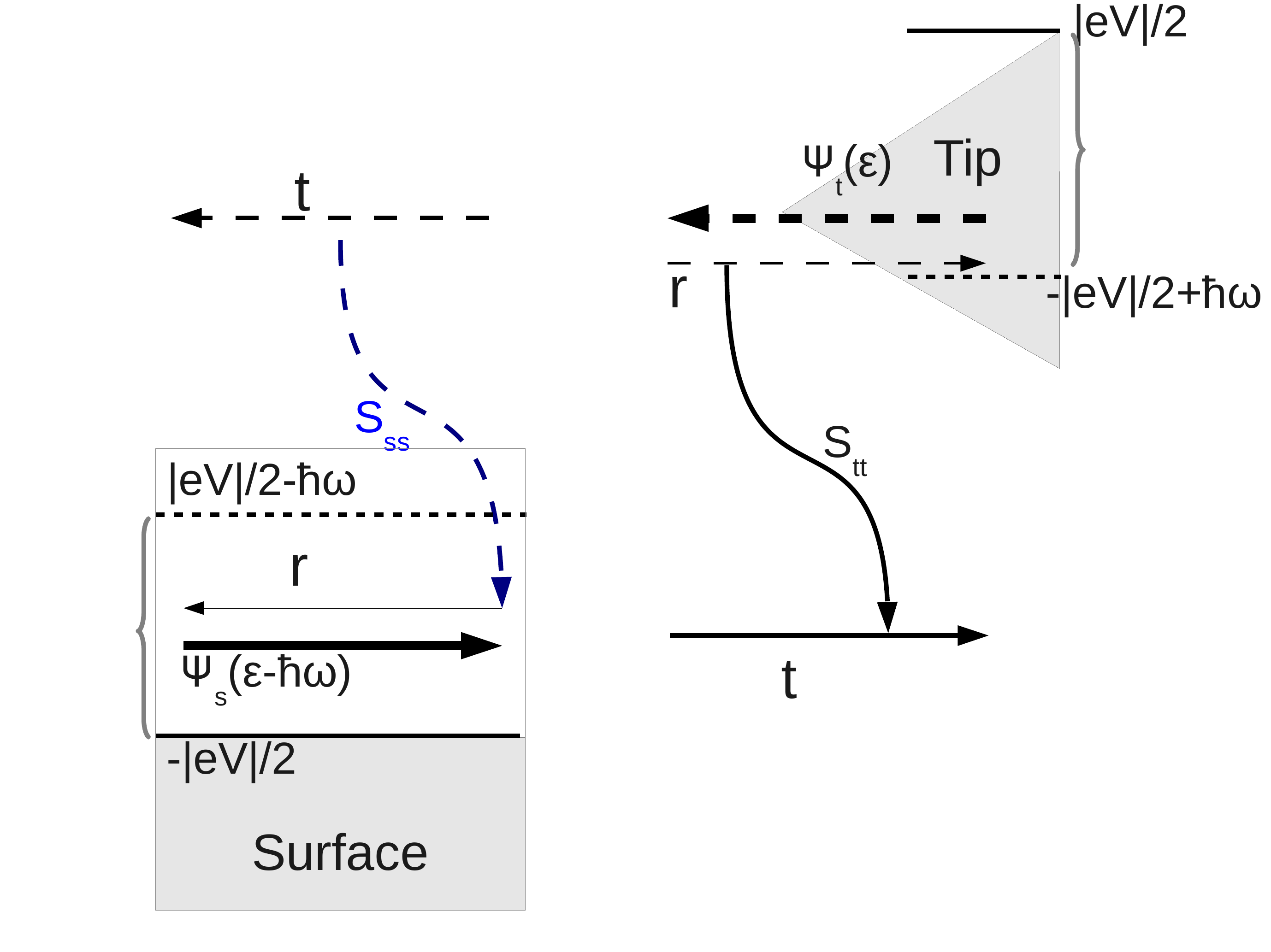}
   \caption{Schematic diagrams showing the two processes contributing to
   $S_{tt}$ (solid black) and $S_{ss}$ (dashed blue) for positive sample bias,
   $V>0$. The curly brackets show two active energy windows for inelastic
   transitions.   
  }
   \label{fig:trandiag}
\end{figure}

Equations~(\ref{eq:arate1-1}-\ref{eq:arate1-2}) show that the finite frequency
noise is related to the eigenchannel transmission and reflection coefficients
at two energy windows. The first energy window corresponds to transmission 
in the energy range $[
\hbar\omega-(eV/2);eV/2]$, the other window is shifted downwards by
$\hbar\omega$, $[-eV/2;eV/2-\hbar\omega]$ . We denote these as the active energy windows. The correlation, $S_{ss}$, corresponds to
inelastic transitions taking place at the device-surface interface.
For positive sample voltage, $V>0$, it is proportional to the transmission coefficient
of the tip scattering state in the high energy window, and the reflection
coefficient of the surface scattering state in the low energy window. The reverse is the case for $S_{tt}$.  Schematic diagrams of these two processes are shown in
Fig.~\ref{fig:trandiag}. 

\section{Results}
Now we apply the method outlined above to calculate the light emission from the
STM resembling two recent experiments where the tip is brought into contact
with (i)a Ag adatom on a Ag(111) surface\cite{ScScBe10}, and (ii) a \buck
molecule a Cu(111) surface\cite{ScLuBrBe12}.  In the experiments, two type of
photons with energy smaller and larger than the applied bias are detected. They
are attributed to one- and two-electron process, respectively.  Here, we focus
only on the former.  We used the SIESTA/TranSIESTA code
\cite{SoArGa.02,BrMoOr.02} with the generalized gradient approximation
(GGA-PBE) for exchange and correlation \cite{PeBuEr.96}. For the Ag-system, we
use a single-$\zeta$ polarized basis-set for the Ag atoms.  For the
\buck-system, we use a double-$\zeta$ basis-set for the carbon atoms, and a
single-$\zeta$ basis-set for the bulk electrode Cu atoms. For both systems, to
accurately describe the surface and/or the chemical bonding with the \buck, an
optimized diffuse basis set was applied for surface layer atoms and the tip
\cite{GaGaLo.09}. 

\subsection{Ag adatom on Ag(111)}
In Ref.~\onlinecite{ScScBe10}, STM-induced light emission from a Ag-Ag(111) junction has
been probed from tunneling to contact regime. The photon yield (roughly emission probability
per electron) develops a plateau in the tunneling regime, and has a kink near
the conductance quantum upon contact. These results suggest possible
correlation between photon emission and current shot noise.

\begin{figure}[htbp]
   \includegraphics[scale=0.35]{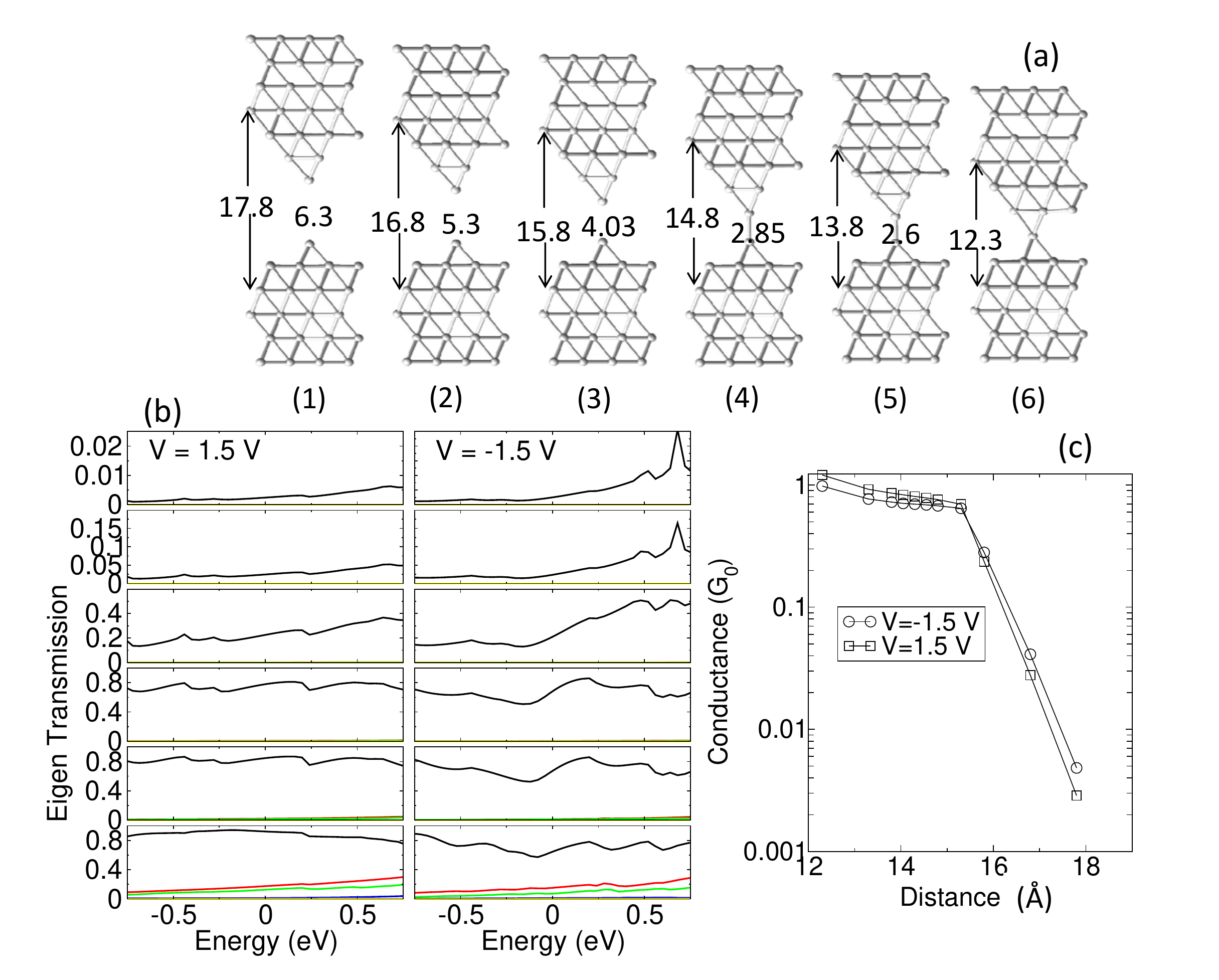}
   \caption{(a) A subset of structures used in the calculation, going from
   tunneling to contact. In the final structure, one tip atom is pushed aside
   when forming contact.  The two surface layers, the tip and the adatom are
   relaxed at zero bias for each structure. The numbers show the distance
   between the two fixed layers and between the tip-adatom in units of \AA. (b)
   Transmission eigenchannels at $V=V_s-V_t=\pm 1.5$ V, going from tunneling to
   contact (top to bottom), for the structures shown in (a). (c) The average conductance as a function of surface
   layer separation, showing the transition from tunneling to contact.}
   \label{fig:AgStruTran}
\end{figure}

To simulate this experiment, we have studied a similar setup: Ag adatom on
Ag(111) surface. Figure~\ref{fig:AgStruTran}(a) shows a subset of the
structures used in the calculations, going from tunneling to contact regime.  A
$4\times 4$ surface unit-cell were used, together with $2\times2$/$5\times5$
surface k-points to sample electronic structure/transmission. We relaxed the
two surface layers, the tip and the adatom at zero bias. After the relaxation,
transport calculations were done for a bias of $V=\pm 1.5$ V.
Figure~\ref{fig:AgStruTran}(b) shows the transmission eigenchannels for the
structures in Fig.~\ref{fig:AgStruTran}(a). From Fig.~\ref{fig:AgStruTran}(b)
it is evident that, (i) there is only one dominate transmission eigenchannel,
and (ii) there is a small asymmetry in the transmission for the two bias
polarities. Figure~\ref{fig:AgStruTran}(c) shows the change of the average
conductance when going from tunneling to contact on a log-scale.  In the
tunneling regime, the conductance depends exponentially on the tip-atom
distance, while it develops to a plateau upon contact as typically seen in
experiments\cite{ScScBe10}.

\begin{figure}[htbp]
   \includegraphics[scale=0.8]{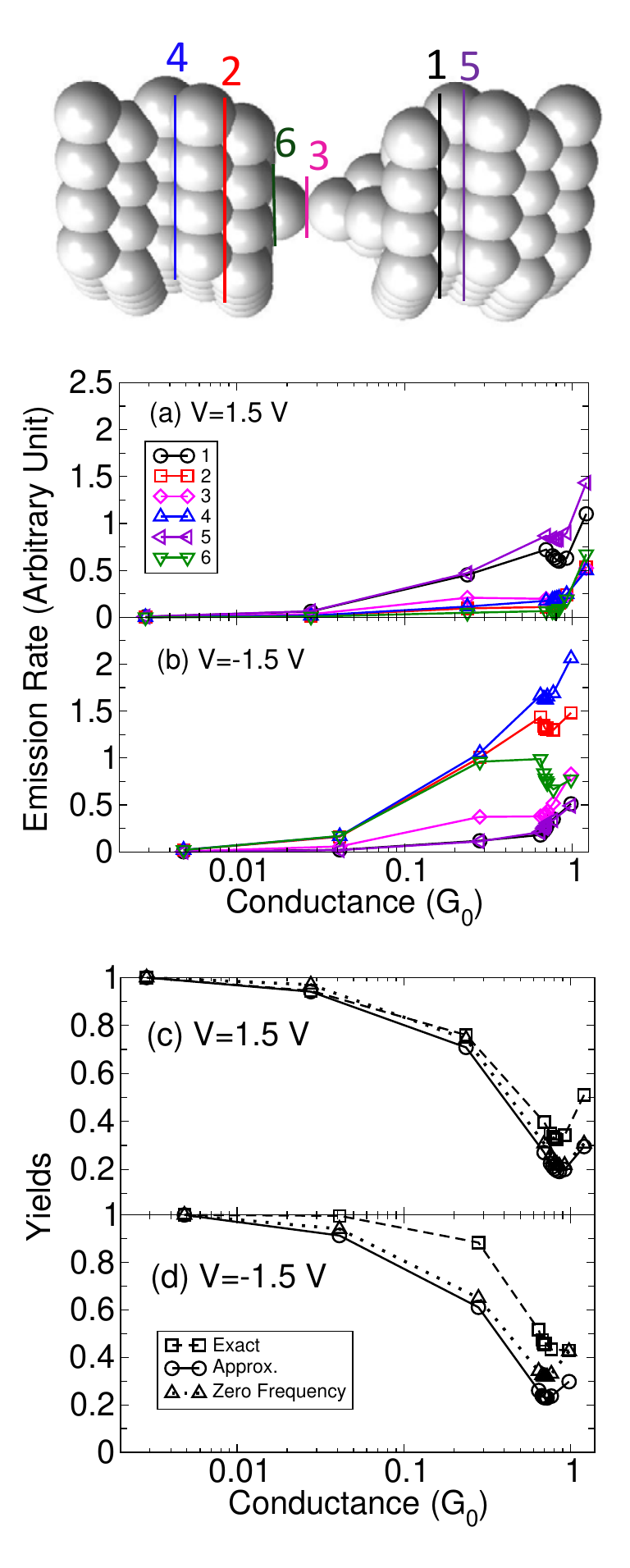}
   \caption{(a)-(b): Calculated noise power (or emission rate) $S_{zz}$ from
   Eq.~(\ref{eq:rate}) for $I_z$ defined through $6$ different surfaces, shown
   above, for plasmon energy $\hbar\Omega=1.2$ eV. (c)-(d): Calculated yields
   $Y=P/\langle I \rangle$, normalized with respect to the first point. The
   power $P$ is the averaged noise power over the $6$ different surfaces (squares).
   Also shown are the results from average of $S_{ss}$ and $S_{tt}$ using the
   approximated expressions Eqs.~(\ref{eq:arate1-1}-\ref{eq:arate1-2})
   (circles), and from the zero-frequency noise calculation used in
   Ref.~\onlinecite{ScScBe10}(triangles). All of them give qualitatively
   similar results.}
   \label{fig:AgEmYi}
\end{figure}

The emission rate (proportional to the shot noise power) was evaluated for a
plasmon energy of $\hbar\Omega=1.2$ eV using Eq.~(\ref{eq:rate}), or
equivalently Eqs.~(\ref{eq:ratenegf}-\ref{eq:ratenegf2}). In order to map out
the spatial distribution, the emission rate were calculated for the surface
current defined at 6 different interfaces, shown in
Fig.~\ref{fig:AgEmYi}(a)-(b). From these calculations, we observe that the
emission rate does not change significantly for interfaces in the same
electrode, while they are quite different for the two electrodes, and for the
tip-adatom interface.  

To relate the emission rates to the intensity of light
emission, we need to do an average of the surface currents, taking into account
the spatial distribution of the plasmon mode, $\xi(z)$.  Since we do not have
specific knowledge about the mode we will choose to do it in the simplest
possible way here. Firstly, we take the equally-weighted average of all the
surface layers (e.g., $\xi(z)=$ Constant). Secondly, as mentioned above, we will
use Eq.~(\ref{eq:fgr2}) instead of Eq.~(\ref{eq:fgr}), so we ignore the cross
terms involving surface current at different positions. 

We have two comments regarding the approximations: (i) In reality, the plasmon
field distribution may change with the tip-surface distance. In the tunneling
regime, we expect a high weighting-factor in the region between the tip-surface
gap.  On the other hand, upon contact, due to the high conductance, we expect
the field distribution to spread out into both
electrodes\cite{scholl2013,Savage2012}. Study of this distance-dependent field
distribution is an interesting problem by itself, and is beyond the scope of
present paper. (ii) We actually tried to include some of the cross terms using
Eq.~(\ref{eq:ratenegf3}), and only see slight change of the final results.
But it is computationally too expensive to include all of them.

The final results for the photon yields $Y=P/\langle I\rangle$, normalized over
the first point, for the two bias polarities are shown in
Fig.~\ref{fig:AgEmYi}(c)-(d). Here the power $P$ is proportional to the
emission rate averaged over six different surfaces. $\langle I \rangle$ is the
average current. In Fig.~\ref{fig:AgEmYi}(c)-(d) , we also show results from the approximate
calculation using Eqs.~(\ref{eq:arate1-1}-\ref{eq:arate1-2}), and from the
zero-frequency noise employed in Ref.~\onlinecite{ScScBe10}. We see that the
qualitatively trends are similar for all these calculations: A plateau
in the tunneling regime, and the development of a dip at contact around the
fully transmitting single channel for $G=1 G_0$, consistent with the
experiments\cite{ScScBe10}.

The agreement between different approximations can be understood from the
eigentransmission plotted in Fig.~\ref{fig:AgStruTran} (b): (i) In the
tunneling regime, there is only one eigenchannel. The eigentransmission is
rather small and scales logarithmically with the distance in the whole energy
range. Consequently, the distance dependence of the photon yields is encoded in
the reflection coefficient $R=1-T\approx 1$. As a result, the photon yields
show a rather weak dependence on the distance. (ii) In the contact regime, the
eigentransmission is rather flat in the whole bias window. From
Eqs.~(\ref{eq:arate1-1}-\ref{eq:arate1-2}), we expect that the finite frequency
shot noise shows weak position dependence, and becomes similar to the zero
frequency one.

\subsection{\buck on Cu(111)}
In Ref.~\onlinecite{ScLuBrBe12}, STM-induced light emission from a \buck
molecule sitting on the reconstructed Cu(111) surface was studied in the
tunneling and contact regime. It was found that the \buck molecule modifies the
photon yields drastically. Especially, a strong bias polarity dependence is
observed, indicating the effect of localized molecular resonance on the light
emission property.

\begin{figure}[htbp]
   \includegraphics[scale=0.6]{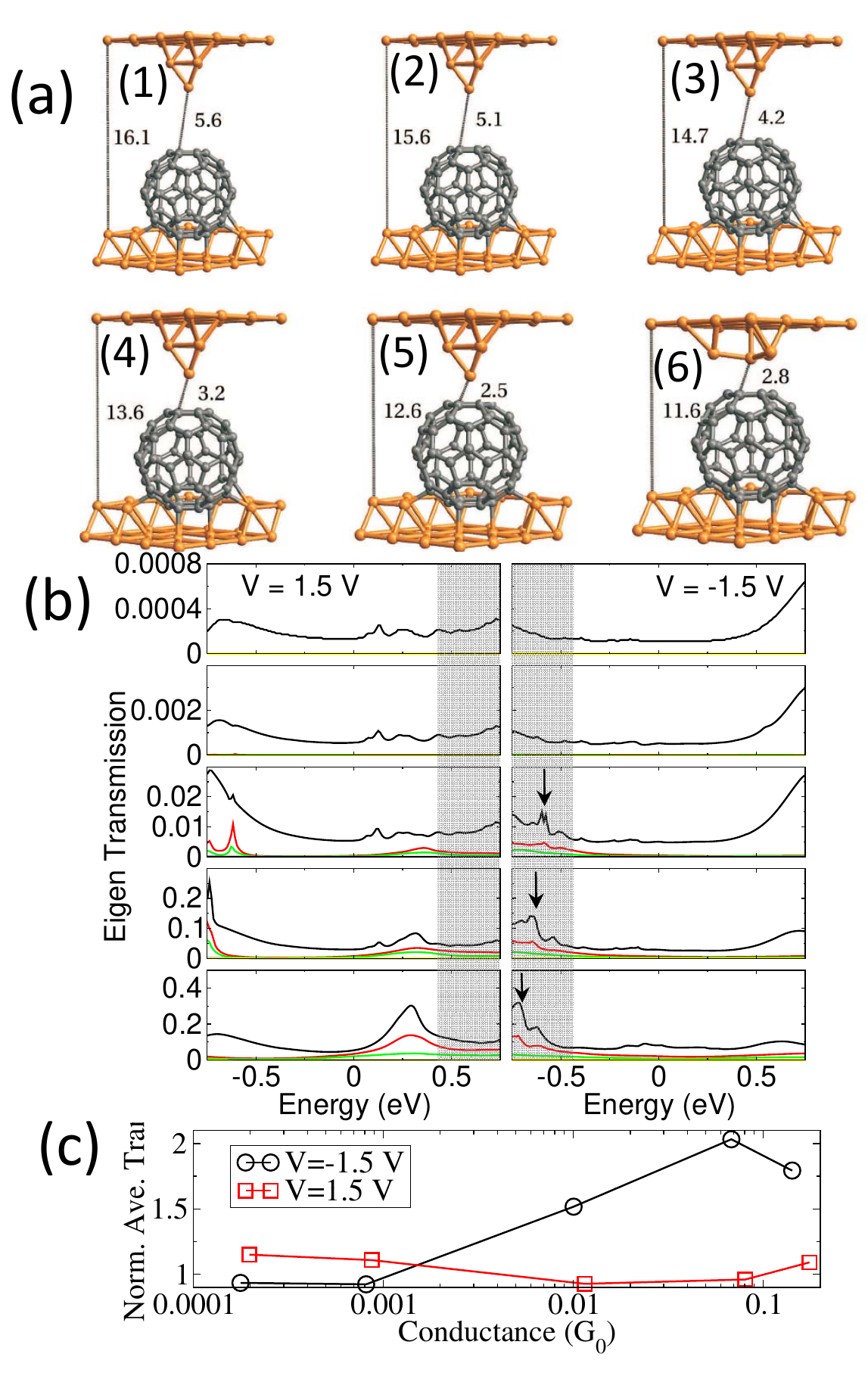}
   \caption{(a) All structures considered in the calculation. In structure 6 a
   deformation of the tip occurred and has been disregarded in the following. 
   The two surface layers, \buck and the tip were relaxed at zero
   bias for each electrode separation. (b) Transmission eigenchannels at $V=\pm
   1.5$ V for the structures shown above. The shaded areas are the active
   energy windows contributing to $S_{ss}$. (c) The
   average transmission in the active energy window (shaded areas in (b)),
   normalized over that in the whole bias window [-0.75 - 0.75] eV. The
   increase from tunneling to contact at $V=-1.5$ V is due to the appearance of
   HOMO level (peak in the shaded region).}
   \label{fig:C60StruTran}
\end{figure}

To simulate this experiment, we used a $4\times 4$ surface
unit-cell, and $2\times2$/$10\times10$ surface k-points in order to sample the
electronic structure/transmission.  Due to the surface reconstruction in the
experiments\cite{Pai2010,ScLuBrBe12} the two first surface layers and tip were
relaxed at zero bias to 0.02~eV/{\AA} at different tip positions. Thus, we do
not capture the abrupt jump-to-contact observed in the experiment at finite
negative bias in our calculations. Figure \ref{fig:C60StruTran} shows the five
different structures considered in the calculations, together with the
transmission eigenchannels at $V=\pm 1.5$ V.  Different from the Ag system,
when making the contact, there are now mainly three contributing eigenchannels.

\begin{figure}[htbp]
   \includegraphics[scale=0.8]{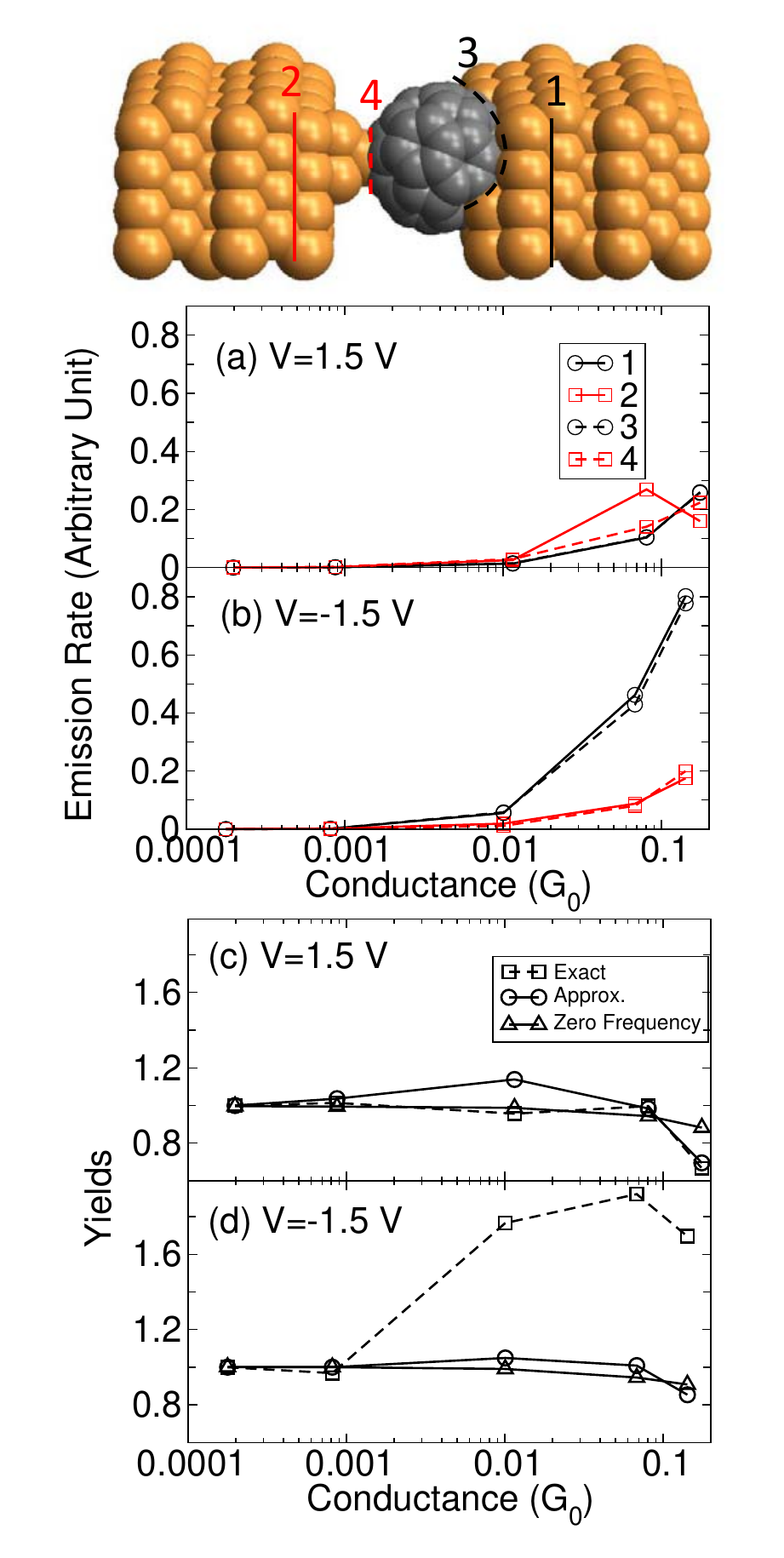}
   \caption{(a)-(b) Similar to Fig.~\ref{fig:AgEmYi}(a)-(b), calculated
   emission rates at $4$ different surfaces for the \buck system using
   $\hbar\Omega=1.2$ eV at $V=\pm 1.5$ V. (c)-(d) Similar to
   Fig.~\ref{fig:AgEmYi} (c)-(d).}
   \label{fig:C60EmYi}
\end{figure}

As in the experiment, we observe different emission rates for the two bias
polarities (Fig.~\ref{fig:C60EmYi}(a)-(b)). For positive sample bias, the
magnitude at 4 different surfaces is comparable. But for the negative bias, the
fluctuations near the surface electrode are $4$ times larger than that of the
tip electrode. Consequently, the calculated yields show different trends at
negative and positive bias when going from tunneling to contact, as shown in
Fig.~\ref{fig:C60EmYi}(c)-(d). These results can be explained as a consequence of the
appearance of the HOMO level in the bias window, as discussed in
Ref.~\onlinecite{ScLuBrBe12}. When the HOMO level enters the bias window, the
occupied charge begins to fluctuate. This generates new available final states
for inelastic transitions, which contribute to high frequency noise at the
plasmon frequency. Since the molecule couples better to the surface than the
tip, the charge fluctuations are compensated mainly by the surface-current
fluctuations. This allows us to understand the results qualitatively by looking at the
surface current fluctuations. In the single channel, small transmission case,
we can ignore the $1-T$ term in Eqs.~(\ref{eq:arate1-1}-\ref{eq:arate1-2}). So
the photon yield due to surface current fluctuation can be characterised by the
ratio of the average transmission in the active window (shaded region in
Fig.~\ref{fig:C60StruTran}) to that in the whole bias window.  We plotted this
normalized average transmission in Fig.~\ref{fig:C60StruTran} (c), and observed
a sudden increase upon contact.

Comparing the two systems, we can see that the main difference between them is
whether spatially localized molecular resonance participates in the light emission
process or not: (1) For the Ag system there are no such localized resonances
and the transmission spectrum is weakly energy dependent. The behavior of the
finite frequency noise is similar to that at zero-frequency. So the
experimental results can basically be understood by looking at the
zero-frequency noise, as has been done in Ref.~\onlinecite{ScScBe10}. (2) On
the other hand for the \buck system, at negative bias, the \buck-HOMO level enters
into the active window upon contact, modifies the transmission in there, and
enhances the shot noise power. From this study, we can see that molecular level
engineering provides an efficient way to control the light emission property of
STM junctions. Along these lines we note that very recent STM experiments using the photon-map technique 
indicate that individual molecular resoances can play a determining role ("gate") for the emission process\cite{LuGrDe.13}.

\section{Conclusions}
We have developed a practical scheme to calculate the finite-frequency shot
noise of the electrical current through a coherent molecular conductor within a
DFT-NEGF approach. By a spatial average, we re-produce qualitatively the
essential features of two recent experiments, confirming the hypothesis that
the current/charge fluctuations are the energy source of STM-induced light
emission from molecular junctions, going from tunneling to contact.
Furthermore, by writing the shot noise expression into a Fermi-golden-rule
form, we have established a connection with the theory of light emission in the
tunneling regime, based on inelastic electronic transitions. The relation
between shot noise power and light emission intensity makes it possible to
understand qualitatively the light emission property of atomic/molecular junctions
with the help of its eigentransmission spectrum.

Here, we have focused on the source of the light emission, which is the
inelastic electronic transitions induced by current. However, to get a
quantitative understanding of the experimental results, in a semi-classical
model of the electron-plasmon coupling, the following questions have to be
addressed: (1) the spatial field distribution of different plasmon modes near
the STM tip, (2) their detailed coupling with the current. These questions are
also important if we want to distinguish the localized gap mode from the
propagating surface mode. Recent experiments showed that the tunneling
electrons can couple to both types. An alternative way to proceed is to perform
time dependent DFT calculations. So far, model structures have been
considered\cite{SoNoGa12} with this approach. However, it is very challenging
to perform calculations on realistic structures involving coupling to the
metallic surfaces in order to approach the experiments.

\appendix
\section{Frequency dependent noise at finite temperature}
\label{app.A}
At finite temperature, to evaluate the surface current correlation, we need all
the matrix elements.  The other three read
\begin{eqnarray*}
	\langle \psi_{s}(\varepsilon)|I_s|\psi_{t}(\varepsilon_-)\rangle &=& -\frac{ie}{\hbar} \langle \psi_{s}(\varepsilon)|W_s^\dagger(\varepsilon,\varepsilon_-)|\psi_{t}(\varepsilon_-)\rangle,\\
	\langle \psi_{t}(\varepsilon)|I_s|\psi_{t}(\varepsilon_-)\rangle &=& \frac{ie}{\hbar} \langle \psi_{t}(\varepsilon)|\Sigma_s(\varepsilon_-)-\Sigma_s^\dagger(\varepsilon)|\psi_{t}(\varepsilon_-)\rangle,\\
	\langle \psi_{s}(\varepsilon)|I_s|\psi_{s}(\varepsilon_-)\rangle &=&\frac{ie}{\hbar} \langle \psi_{s}(\varepsilon)|\Sigma_t^\dagger(\varepsilon)-\Sigma_t(\varepsilon_-)-\omega I|\psi_{s}(\varepsilon_-)\rangle.
\end{eqnarray*}
Assuming a constant self-energy, for positive sample bias, we have the
full result for surface current noise at finite temperature
\begin{eqnarray}
	S_{ss}(\omega) &=& \frac{e^2}{2\pi\hbar}\sum_{\alpha\beta}C_{\alpha\beta}(\omega)\Delta n_F^{\alpha\beta} ,\nonumber
\end{eqnarray}
with
\begin{eqnarray}
	C_{tt}(\omega)&=&\int {\rm Tr}\left[ T(\varepsilon)T(\varepsilon_-) \right]\Delta n_F^{tt}d\varepsilon,\nonumber\\
	C_{ss}(\omega)&=&\int {\rm Tr}\left[ (\omega I -i\Gamma_t)A_s(\varepsilon_-)(\omega I+i\Gamma_t)A_s(\varepsilon) \right]\Delta n_F^{ss}d\varepsilon,\nonumber\\
	C_{st}&=&\int {\rm Tr}\left[ (I-T(\varepsilon))T(\varepsilon_-) \right]\Delta n_F^{ts}d\varepsilon,\nonumber\\
	C_{ts}&=&\int {\rm Tr}\left[ (I-T(\varepsilon_-))T(\varepsilon) \right]\Delta n_F^{ts}d\varepsilon,\nonumber
	\label{}
\end{eqnarray}
where 
\begin{equation}
	\Delta n_F^{\alpha\beta} = n_F(\varepsilon,\mu_\alpha)(1-n_F(\varepsilon_-,\mu_\beta))\nonumber.
	\label{}
\end{equation}
The above result includes both the Nyquist-Johnson (thermal) and the shot
noise.  Notice the different form of $C_{ss}$ from $C_{tt}$. It is related to
the complex reflection coefficients in the scattering approach discussed by
B\"uttiker\cite{But92}. Physically, it means that even when the transmission
is zero, there still could be fluctuations at the surface electrode at finite
temperature.

\section*{Acknowledgements}
We thank Prof. R. Berndt and Dr. N. Schneider for insightful discussions, and the Danish
Center for Scientific Computing(DCSC) for providing computer resources. J. T. L\"u is supported
by the Fundamental Research Funds for the Central Universities, HUST:2013TS032.

\bibliography{bibtheory}

\end{document}